%% file: pi0U.tex
\documentclass[final]{elsarticle}
\usepackage{epsfig}
\usepackage{amssymb}
\usepackage{amsmath}
\graphicspath{{fig/}}

\newcommand{\E}[1]{\mbox{$\times$10$^{#1}$}}
\newcommand{\EA}[1]{\mbox{10$^{#1}$}}

\newcommand{\po}{\mbox{$\pi^0$}}

\begin{document}

\begin{frontmatter}

\title{Search for a dark photon in the $\pi^0\to e^+e^-\gamma$ decay}

\input{author1.tex}


\begin{abstract}

The  presently  world largest  data  sample  for $\po\to\gamma  e^+e^-$
decays studies containing  nearly 5\E{5} events  was collected using  the WASA
detector at  COSY.  A  search for  a dark photon  $U$ produced  in the
$\po\to\gamma  U\to\gamma  e^+e^-$  decay  from  the  $pp\to  pp\pi^0$
reaction  was  carried out.   An  upper limit  on  the  square of  the
$U-\gamma$  mixing  strength   parameter  $\epsilon^2$  of  $  5\times
10^{-6}$ at 90\%  CL was obtained for the mass  range 20 MeV $<M_{U}<$
100 MeV.   This result together  with other recent  experimental limits
significantly  reduces  the $M_U$  {\it  vs.}  $\epsilon^2$  parameter
space which could explain the presently seen deviation between the 
Standard Model prediction and the direct measurement of the anomalous 
magnetic moment of the muon. 

\end{abstract}



\end{frontmatter}

\section{Introduction}
\label{Sec:Intro}

Decays of  neutral pseudoscalar  mesons into a  lepton-antilepton pair
and  a photon,  $P\rightarrow l^+l^-\gamma$,  are among  the  processes to
search for a  new light vector boson connected  with dark gauge forces
\cite{Fayet:1980ad,Dobroliubov:1988pe,Boehm:2003hm}.    An extra $U(1)$  
boson  is
postulated  in most extensions of the Standard Model. Recent interest in
searches of a light vector boson, in the ${\cal O}$(MeV--GeV) mass range, 
is motivated by astrophysics observations such as  the positron  and/or electron  excesses
observed by PAMELA \cite{Adriani:2008zr}, ATIC \cite{Chang:2008aa} and
H.E.S.S. \cite{Aharonian:2008aa} as well  as the narrow 0.511 MeV $\gamma$
ray   emission  from   the   galactic  bulge   observed  by   INTEGRAL
\cite{Jean:2003ci}.

\begin{figure}
\centerline{a)\includegraphics[width=0.33\textwidth,clip=]{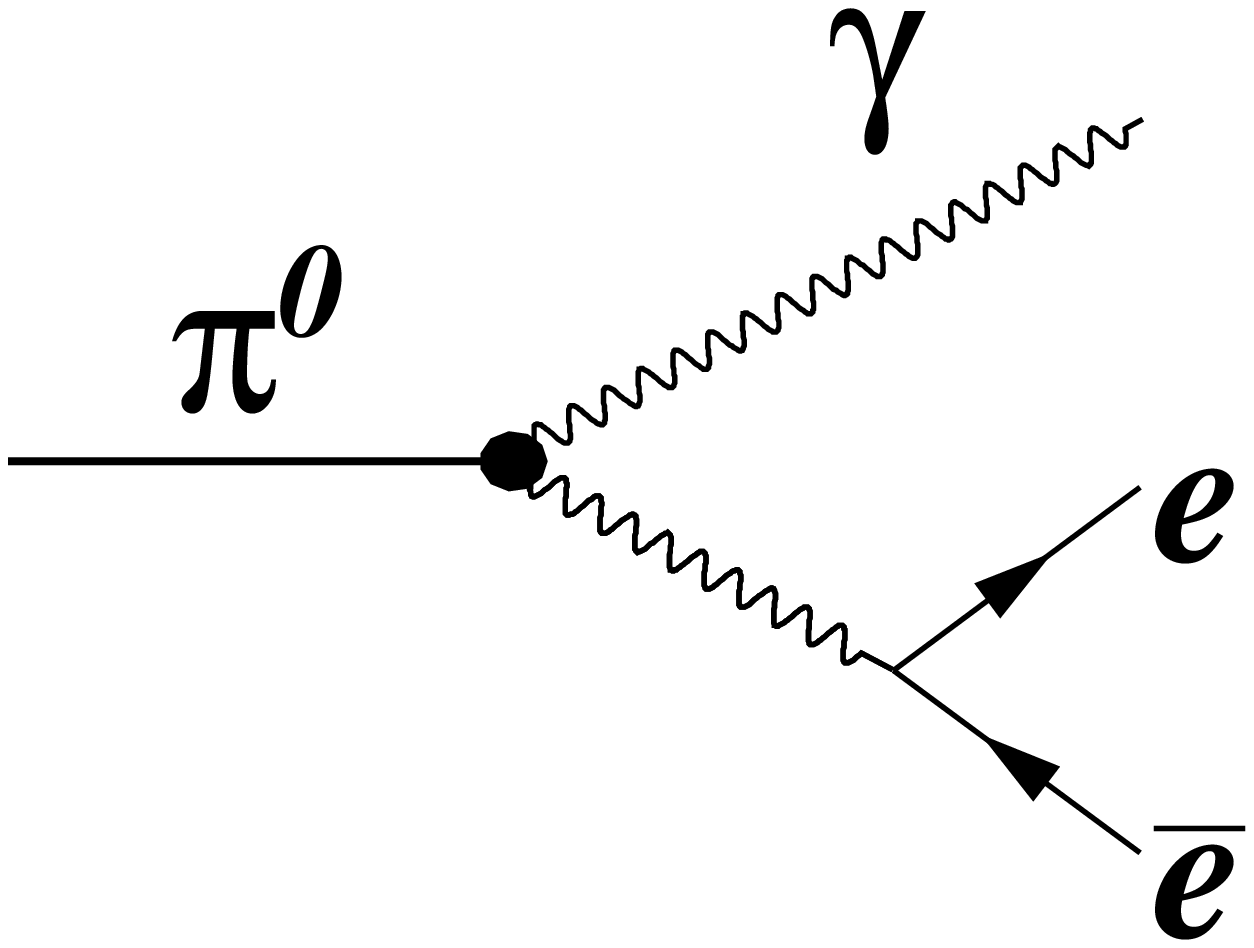}
b)\includegraphics[width=0.33\textwidth,clip=]{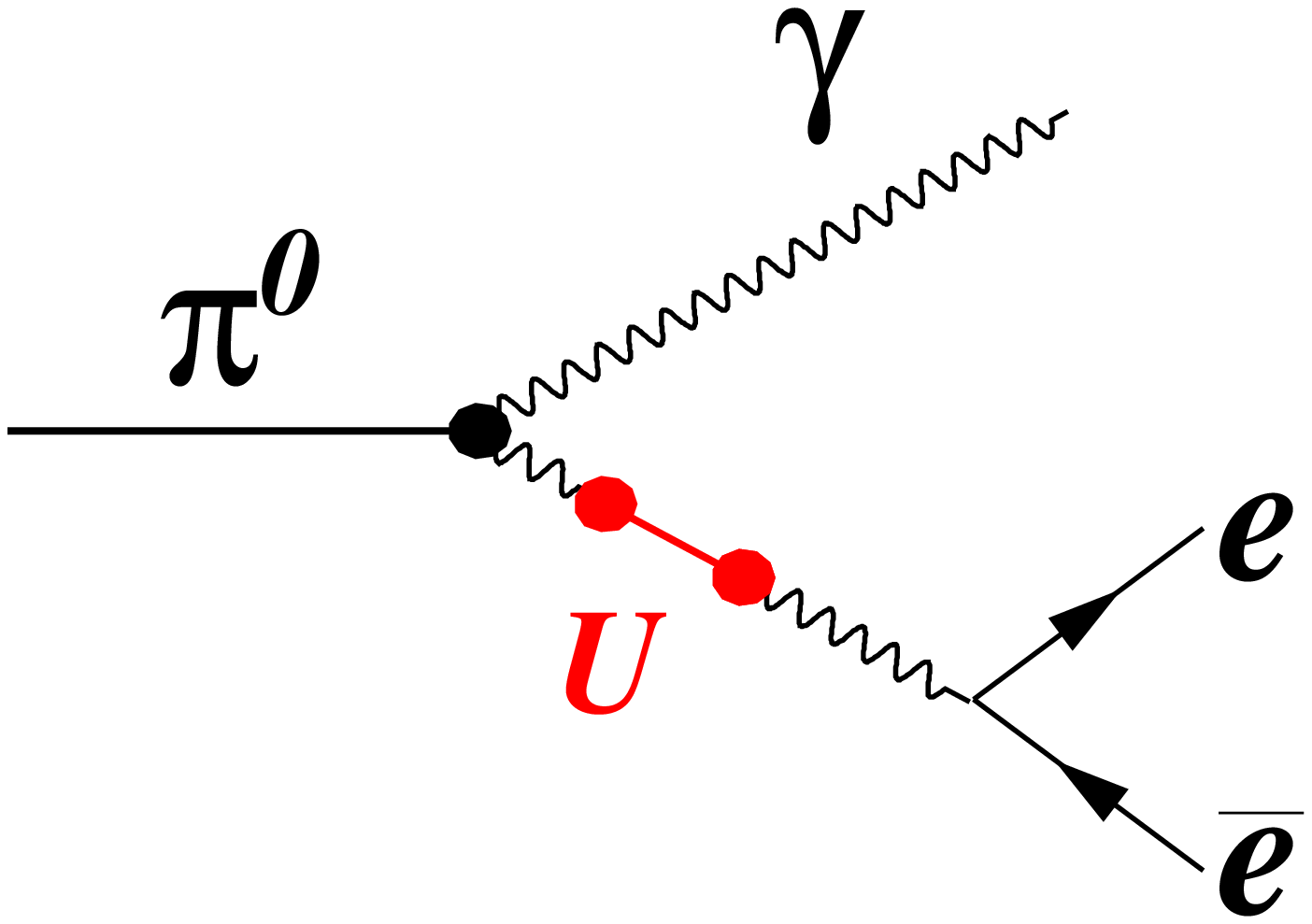}
c)\ \includegraphics[width=0.29\textwidth,angle=-90,origin=br,clip=]{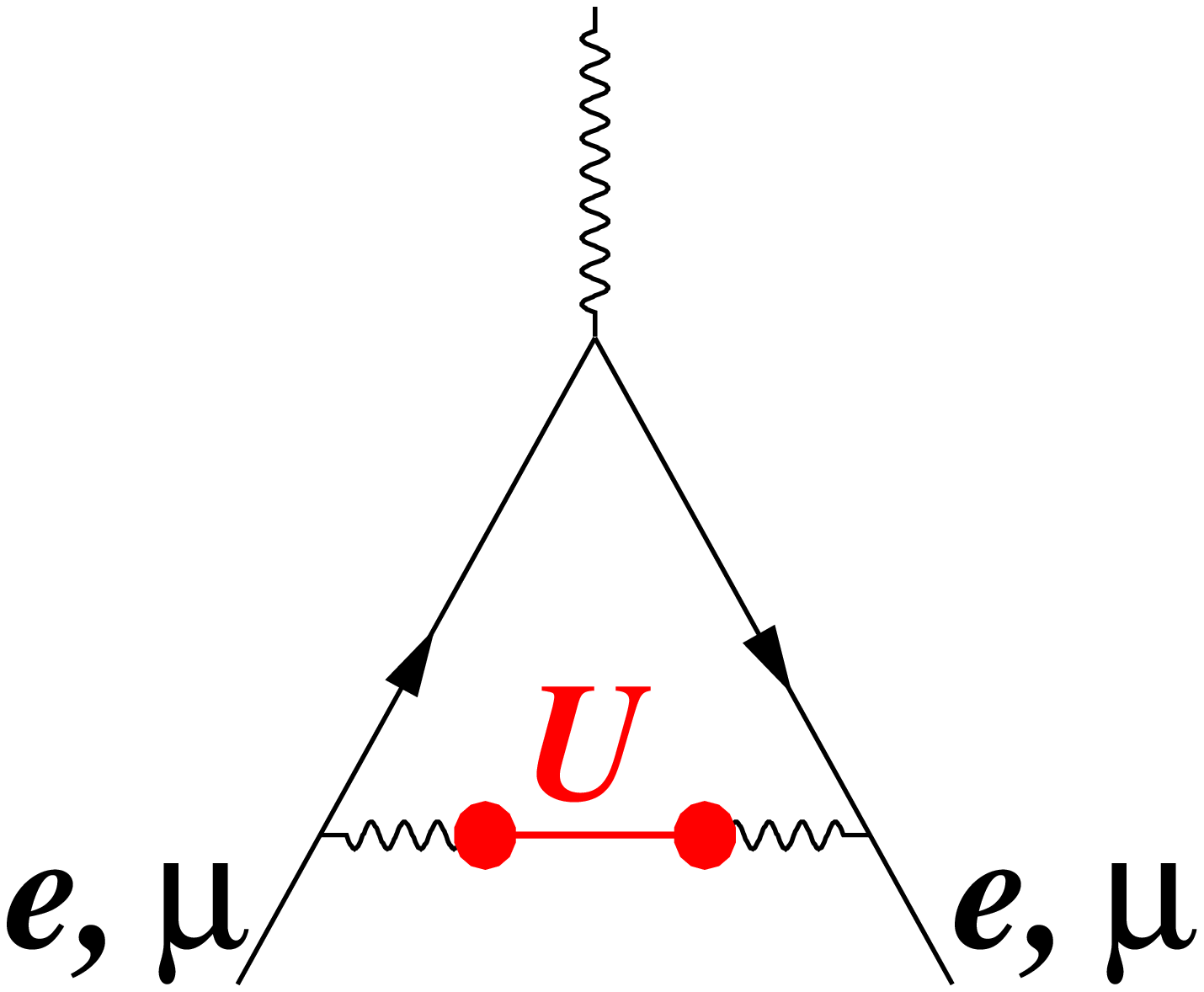}}
\caption[aa]{\label{fig:hee2b} Feynman  diagrams for 
a)  the lowest order electromagnetic $\pi^0\to e^+e^-\gamma$ decay
and a possible  contribution of $U$ vector boson to: b) 
$\pi^0\to e^+e^-\gamma$ and c) lepton $g-2$.}
\end{figure}

In one of the simplest scenarios dark matter particles belonging to an
additional  abelian gauge  symmetry are  added to  the  Standard Model
(SM).    The  new   symmetry   leaves  the   SM  particles   unchanged
\cite{Fayet:1980rr,Holdom:1985ag,Boehm:2003hm,Pospelov:2007mp}.
The associated gauge boson can communicate with the SM through a small
mixing in the kinetic term of the QED Lagrangian \cite{Okun:1982xi}:
\begin{equation}
{\cal L}_{\rm mix} \, = \, -\frac{\epsilon}{2} \, F_{\mu\nu}^{\rm QED} F^{\mu\nu}_{\rm dark} 
\label{eq:fdark}
\end{equation}
where  $\epsilon$ is  the mixing  parameter. The  gauge boson  $U$ (also 
$A'$, $\gamma'$ or $Z_d'$) is  often called a {\em dark photon} since it can  mix with the
photon in all processes (examples are shown in Figs.~\ref{fig:hee2b}b
and         1c).          Phenomenological         arguments
\cite{ArkaniHamed:2008qn,Dienes:1996zr,Goodsell:2009xc}  suggest  that the  $\epsilon$
parameter must be of the order of $10^{-4}-10^{-2}$ and the boson mass
$M_U$  below  2   GeV.   This  estimate  is  also   supported  by  the
astrophysical  observations and the  constraints imposed  by precision
measurements such  as the anomalous  magnetic moments ($g-2$)  of muon
and  electron  \cite{Pospelov:2008zw}.  The  contribution  of the  $U$
boson to the $(g-2)_l$  ($l=e,\mu$) (Fig.~\ref{fig:hee2b}c) is given in \cite{Pospelov:2008zw}
by:
\begin{equation}
  \Delta             (g-2)_l=\frac{\alpha\epsilon^2}{\pi}\int_0^1            dz
  \frac{2m_l^2z(1-z)^2}{m_l^2(1-z)^2+M_U^2z}.
\end{equation}
Investigations  of the  $M_U$ {\it  vs.} $\epsilon^2$  parameter space
corresponding  to  the   experimentally  preferred $(g-2)_\mu$  value
(shifted $+3.6\sigma$  with  respect  to   the  SM  value
\cite{Bennett:2006fi,Davier:2010nc,Hagiwara:2011af}) are therefore of great importance.

For  a $U$ boson
with mass less  than twice the muon mass the total  decay width is for
all  practical purposes (neglecting  higher-order electric,  tiny weak
interaction contributions from the $U$  boson -- $Z_0$ coupling, and the
decay   to  light   dark  scalars   and/or  fermions)    given  by
\cite{Bjorken:2009mm,Batell:2009yf}:
\begin{equation}
\Gamma_U=\Gamma_{ U \to e^+e^- }=\frac{1}{3} \alpha \epsilon^2 \, M_U \, 
\sqrt{1-\frac{4 m_e^2}{M_U^2} }\left(1+\frac{2 m_e^2}{M_U^2}\right), 
\label{eqn:width}
\end{equation}
where $m_e$ is the electron mass.

Such  a  light  $U$  boson   can  be  directly  produced  in  particle
accelerators, see                         {\it                        e.g.}
Refs~\cite{Bjorken:2009mm,Batell:2009yf,Reece:2009un,Freytsis:2009bh,Borodatchenkova:2005ct,Yin:2009mc,Essig:2009nc,Baumgart:2009tn,Li:2009wz}. The
idea is to search for narrow structures in the invariant mass spectrum
of the lepton-antilepton pair.

The $M_U$ {\it vs.}  $\epsilon^2$ region corresponding to the measured
$(g-2)_\mu$ value $\pm 2\sigma$ is  covered by the data from the BABAR
\cite{Aubert:2009cp},    MAMI    A1    \cite{Merkel:2011ze},    KLOE-2
\cite{Archilli:2011zc}  and APEX  \cite{Abrahamyan:2011gv} experiments
for  $M_U$  masses above 100 MeV.  On the  lower end this preferred
region  is   excluded  by  the   $(g-2)_e$  value  for   $M_U<30$  MeV
\cite{Endo:2012hp,Aoyama:2012wj}.  In  addition,  $\epsilon^2$ regions
below  \EA{-12} are  excluded  by experiments  which  are sensitive  to
lepton pairs  from displaced secondary  vertices ($\tau_U>10^{-11}$ s)
\cite{Amsler:1994gt,Altegoer:1998qta,Gninenko:2011uv}.

Our experiment  aims at searching for  a short-lived $U$  boson in the
$\pi^0$  Dalitz  decay, $\pi^0\to  e^+e^-\gamma$,  covering the  range
preferred by  the experimental  value of $(g-2)_\mu$  for $20$ MeV$<M_U<100$
MeV.  In  this region, for  $\epsilon^2 >\EA{-6}$ the  average distance
passed by  a boson emitted from  a low energy $\pi^0$  decay should be
less  than a  millimeter.  The  best limit  from a  previous $\pi^0\to
e^+e^-\gamma$ experiment with the origin of the $e^+e^-$ pair close to
the production  vertex was obtained by the  SINDRUM collaboration more
than twenty  years ago \cite{MeijerDrees:1992kd,Gninenko:2013sr}.  The
SINDRUM result is  based on a sample of  98400 $\pi^0\to e^+e^-\gamma$
decays with $e^+e^-$ invariant masses above 25 MeV.

\section{The Experiment}

The WASA detector setup was built and first used at CELSIUS in Uppsala
and moved to COSY (COoler  SYnchrotron) J\"ulich in the Summer of 2005
\cite{Adam:2004ch}.   The  detector  was  designed and  optimized  for
studies of rare $\pi^0$ meson decays produced in $pp\to pp\pi^0$ reaction
\cite{Bargholtz:2008aa}.      It     consists     of    three     main
components: \newline The Forward  Detector (FD) -- covering scattering
angles in the $3^\circ-18^\circ$ range used for tagging and triggering
of meson production,  the Central Detector (CD) --  used for measuring
meson decay products,  and the pellet target system.   The target beam
consists of 20 -- 30~$\mu$m diameter pellets of hydrogen, providing an
areal  target density  in  the order  of  $10^{15}$ atoms/cm$^2$.  The
diameter of the pellet beam is $\sim$ 3.8 mm.

The CD surrounds the interaction  region and is designed to detect and
identify  photons, electrons, and  charged pions.   It consists  of an
inner drift  chamber (MDC),  a superconducting solenoid  providing the
magnetic field  for momentum determination,  a barrel of  thin plastic
scintillators (PS) for particle  identification and triggering, and an
electromagnetic  calorimeter.  The  amount of  structural  material is
kept  to a  minimum to  reduce  the amount  of secondary  interactions
outside of  the detector sensitive  volumes.  The beryllium  beam pipe
(diameter  6  cm)  wall  is  1.2~mm  thick and  the  material  of  the
superconducting solenoid corresponds to 0.18 radiation lengths.

The FD  allows identification and  reconstruction of protons  from the
$pp\to pp\pi^0$  reaction close  to threshold.  The  track coordinates
are  provided by four  sets of  straw proportional  chambers.  Kinetic
energies are reconstructed using  the $\Delta E$ information in layers
of  plastic scintillators  of different  thickness.  In  addition, the
signals  are used  for triggering.   The kinetic  energy, $T$,  of the
protons can be  reconstructed with a resolution of  $\sigma(T)/ T \sim
1.5-3\%$ for kinetic energies below 400~MeV.

The results presented here are based on data collected during one-week
WASA-at-COSY   run carried out in 2010.   The $\pi^0$ mesons were  produced in
proton--proton interactions at a  kinetic beam energy of 550~MeV.  The
beam energy corresponds to the center-of-mass excess energy of 122~MeV
with  respect  to $pp\pi^0$  threshold  ({\it  i.e.}   below two  pion
production     thresholds)     with     a     cross     section     of
1.12~mb~\cite{Rappenecker:1995pw}.   The maximum  scattering  angle of
the outgoing  protons for the  reaction is 45$^\circ$.   For detection
and for  triggering purposes  the phase space  of the  $pp\to pp\pi^0$
reaction can be divided into three regions:
\begin{enumerate}
\item  Both  protons are  measured  in  the  FD. This  corresponds  to
  a geometrical acceptance of 19\%.
\item One proton is measured in the FD and  one  in the
  forward  part  of  the   PS (scattering  angles
  $20^\circ-40^\circ$).   This corresponds to a geometrical
acceptance of   42\%.
\item  Both  protons  are  registered  in  the PS.   This  corresponds  to
  a geometrical acceptance of 21\%.
\end{enumerate}

Case (1) allows the definition of the most selective trigger condition
and the  best resolution in the  missing mass with respect  to the two
protons. Therefore,  the main trigger for the  experiment required two
tracks in the FD.  The  protons from the $pp\to pp\pi^0$ reaction have
a maximum kinetic energy of 350  MeV and are mostly stopped in the FD.
This allows the inclusion of a veto from a thin plastic detector layer
placed  at the  far end  of  the FD  into the  trigger condition.   In
addition, two  hits in the central  part of the  PS (scattering angles
$45^\circ  -   135^\circ$)  were   required,  aiming  to   select  the
electron-positron pair.  An additional,  scaled down, trigger based on
case  (2) was  used in  parallel.  The  WASA-at-COSY  data acquisition
system allowed  the collection of  more than \EA{4} events  per second
and the  luminosity was  set to optimize  the conditions for  the main
trigger.   The  integrated  luminosity  of  the  run  was  about  0.55
pb$^{-1}$.

\begin{figure}
\begin{center}
a)\includegraphics[width=0.6\textwidth]{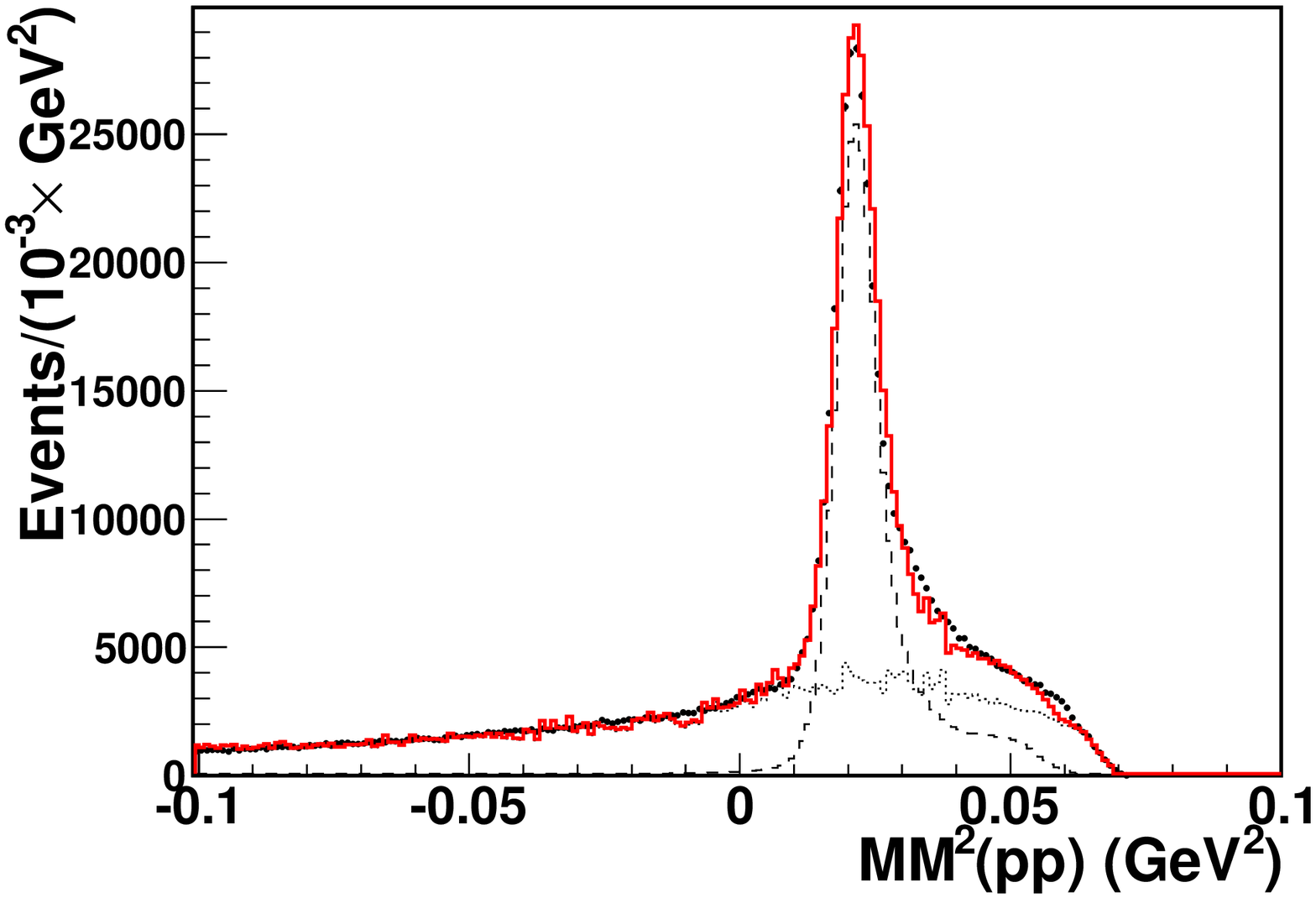}
b)\includegraphics[width=0.6\textwidth]{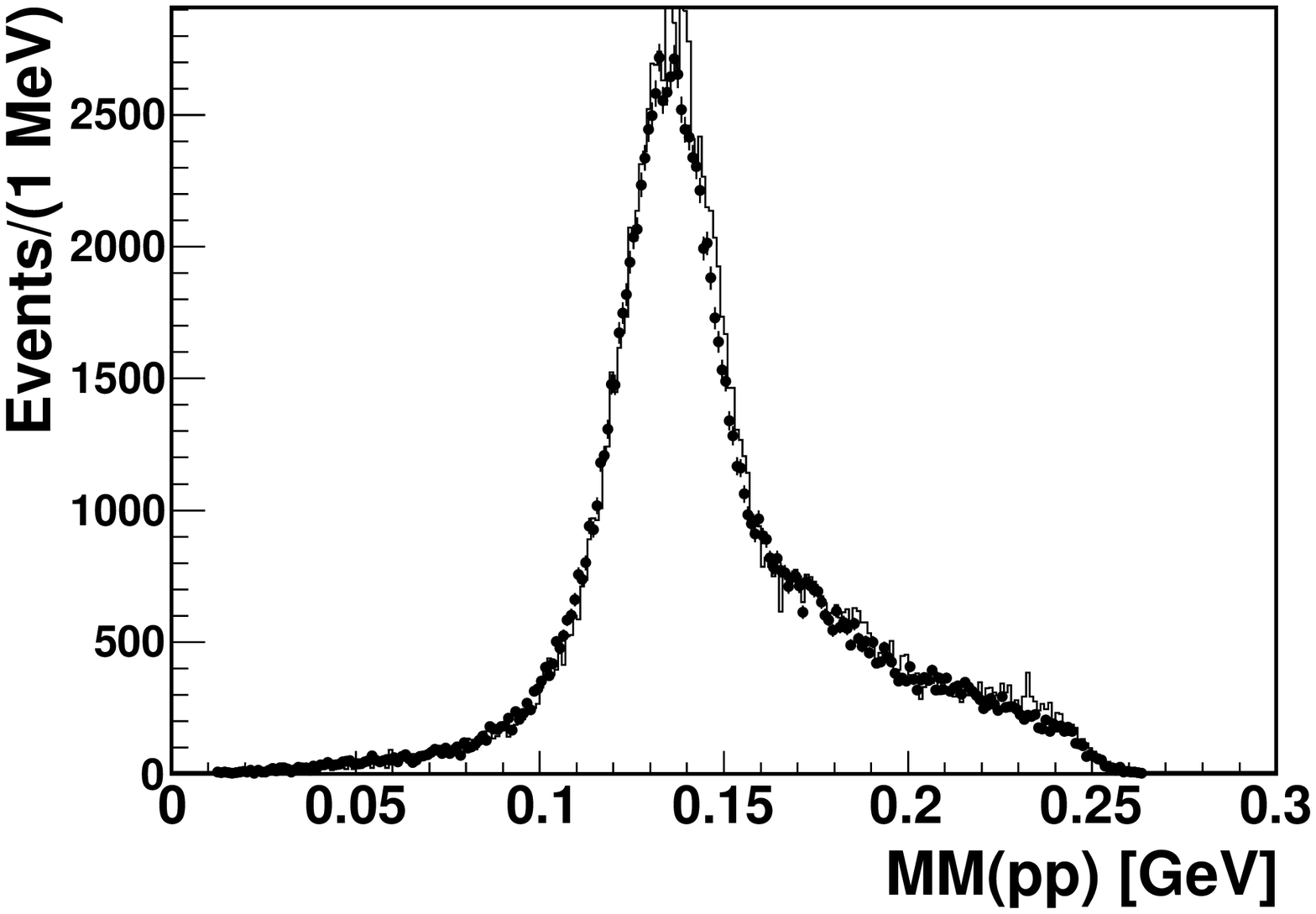}
c)\includegraphics[width=0.6\textwidth]{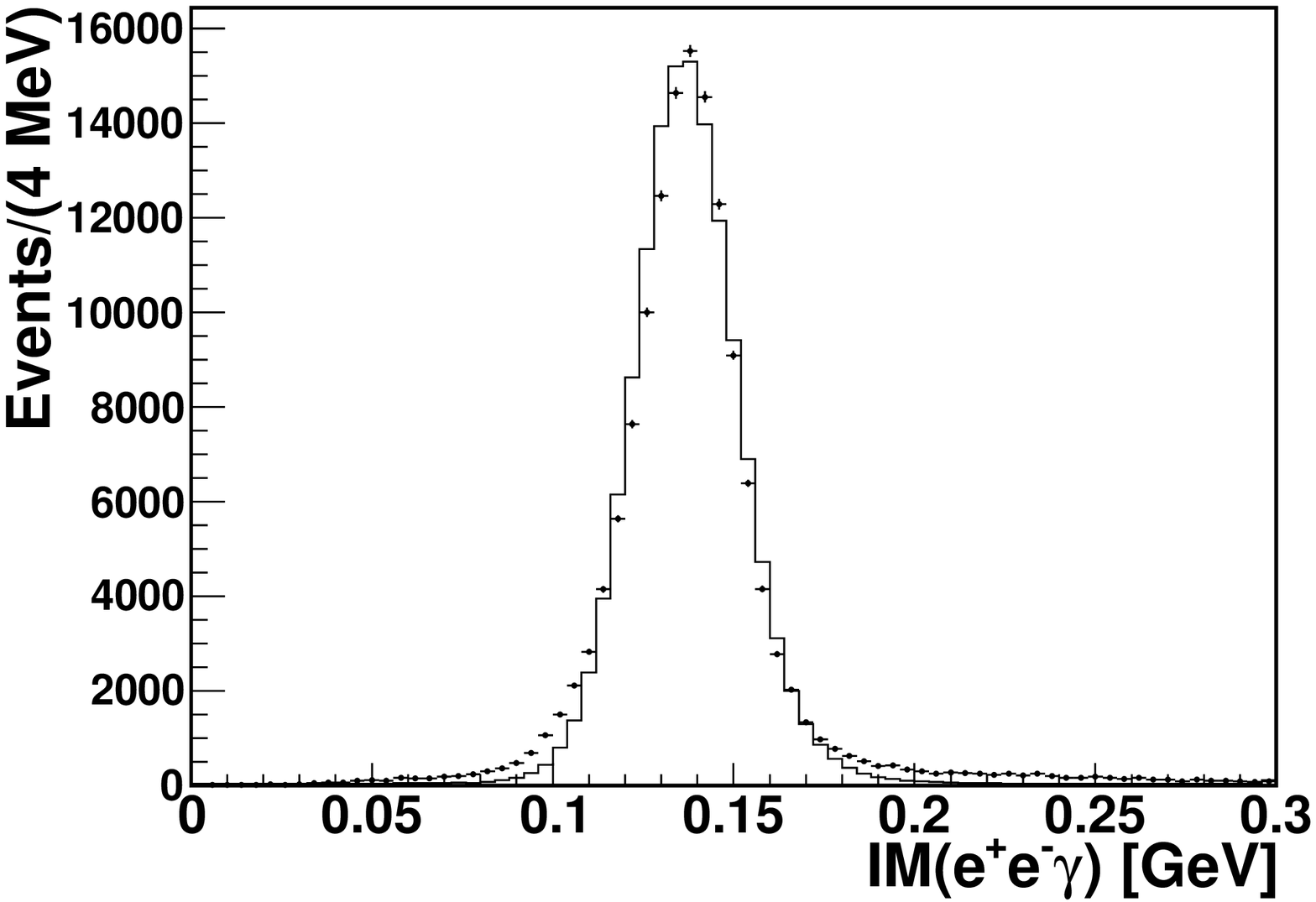}
\end{center}
\caption{\label{fig:mmfd2}Detector performance plots for a data sample  with  two reconstructed protons,  an
  $e^+e^-$ pair  and a  photon.  a) Distribution  of the  missing mass
  squared with respect to the  two protons registered in the FD before
  electron   identification.    Experimental   data  (black   points);
  simulations:  $\pi^0\to   e^+e^-\gamma$  and  $\pi^0\to\gamma\gamma$
  (broken line), random coincidences  of two events (dotted line), and
  the sum  (solid line).  b)  Distribution of $MM(pp)$  after electron
  identification: experimental  data (black  points) and sum  of Monte
  Carlo    simulations   (solid   line).     c)\label{fig:imeeg}   The
  reconstructed  invariant  mass of  the  $e^+e^-\gamma$ system  after
  particle identification  cut.}
\end{figure}

The data quality  is illustrated by analysis of  the main trigger data
sample and  requesting in the  analysis two identified  (using $\Delta
E/\Delta E$  method) FD proton  tracks.  An electron positron  pair is
selected by  requiring two  oppositely curved tracks  in the  MDC with
scattering angles  between $40^\circ$  and $140^\circ$.  A  photon hit
cluster in the calorimeter with an energy deposit above 20 MeV is also
requested.   The missing  mass  squared with  respect  to two  protons
($MM^2(pp)$)    for    the     above    selection    is    shown    in
Fig.~\ref{fig:mmfd2}a.   In addition to  the $pp\to  pp\pi^0$ reaction
signal  one sees  also a  contribution due  to random  coincidences of
$pp\to  pp$  and  $pp\to   pn\pi^+$  reactions.   This  background  is
effectively   suppressed   by    including   electron   and   positron
identification using the reconstructed momentum and the energy deposit
in  the  calorimeter.   The  corresponding
$MM(pp)$ plot  after this cut is shown  in Fig.~\ref{fig:mmfd2}b.  The
$\pi^0\to  e^+e^-\gamma$ decay  is independently  identified  from the
invariant  mass  of the  decay  products  $IM(e^+e^-\gamma)$ (calculated 
assuming the tracks originate at the beam target crossing) shown  in
Fig.~\ref{fig:imeeg}c.  The data are well described by a simulation of
$pp\to   pp\pi^0$   with   $\pi^0\to   e^+e^-\gamma$   and   $\pi^0\to
\gamma\gamma$ decays, where in the  latter case one of the two photons
converts in the beryllium beam tube.

For  Monte Carlo  simulations,  angular distributions  for the  $pp\to
pp\pi^0$  reaction from  ~\cite{Rappenecker:1995pw} were  used  in the
event  generation.  The  $\pi^0\to e^+e^-  \gamma$ decay  is generated
using the lowest order QED matrix element squared:
\begin{equation}
|\mathcal A|^2=\Gamma_{\gamma\gamma} 16\pi^3 M
\frac{\alpha}{\pi}\frac{1}{q^2} \left(1-\frac{q^2}{M^2}\right)^2
\left(1+\cos^2\theta^*+\frac{4 m_e^2}{q^2}\sin^2\theta^*
\right)|F(q^2)|^2 \label{matreeg}
\end{equation}
where $\theta^*$ is the angle of $e^+$ in the dilepton rest frame with
respect to the dilepton momentum  in the overall $\pi^0$ decay system,
$M$   and  $m_e$   are  $\pi^0$   and  $e^\pm$   masses  respectively,
$\Gamma_{\gamma\gamma}$ is  the partial $\pi^0\to  \gamma\gamma$ decay
width, and $F(q^2)$  (with $q^2$ the squared momentum  transfer of the
off-shell  photon) is the  $\pi^0$ transition  form factor.   The form
factor close to $q^2=0$  is parametrized as: $F(q^2)=1+aq^2/M^2$.  The
value  of  the  dimensionless  linear  coefficient  $a$  is  $0.032\pm
0.004$~\cite{Beringer:1900zz}.

The  matrix element  from  Eqn.~(\ref{matreeg}) leads  to the  following
unperturbed $d\Gamma/dq$ distribution~\cite{Landsberg:1986fd}
for the standard lowest order electromagnetic decay
   $\pi^0 \to e^+ e^- \gamma$ of Fig.~\ref{fig:hee2b}a:
\begin{equation}
\frac{d\Gamma}{dq} = \Gamma_{\gamma\gamma}
\frac{4\alpha}{3\pi}\frac{1}{q}\sqrt{1-\frac{4m_e^2}{q^2}}
\left(1+\frac{2m_e^2}{q^2}\right)
\left(1-\frac{q^2}{M^2}\right)^3|F(q^2)|^2.
\label{eqn:dgdq2}
\end{equation}

\section{Data analysis}
\label{sec:analysis}
\begin{figure}
a)\includegraphics[width=0.98\textwidth]{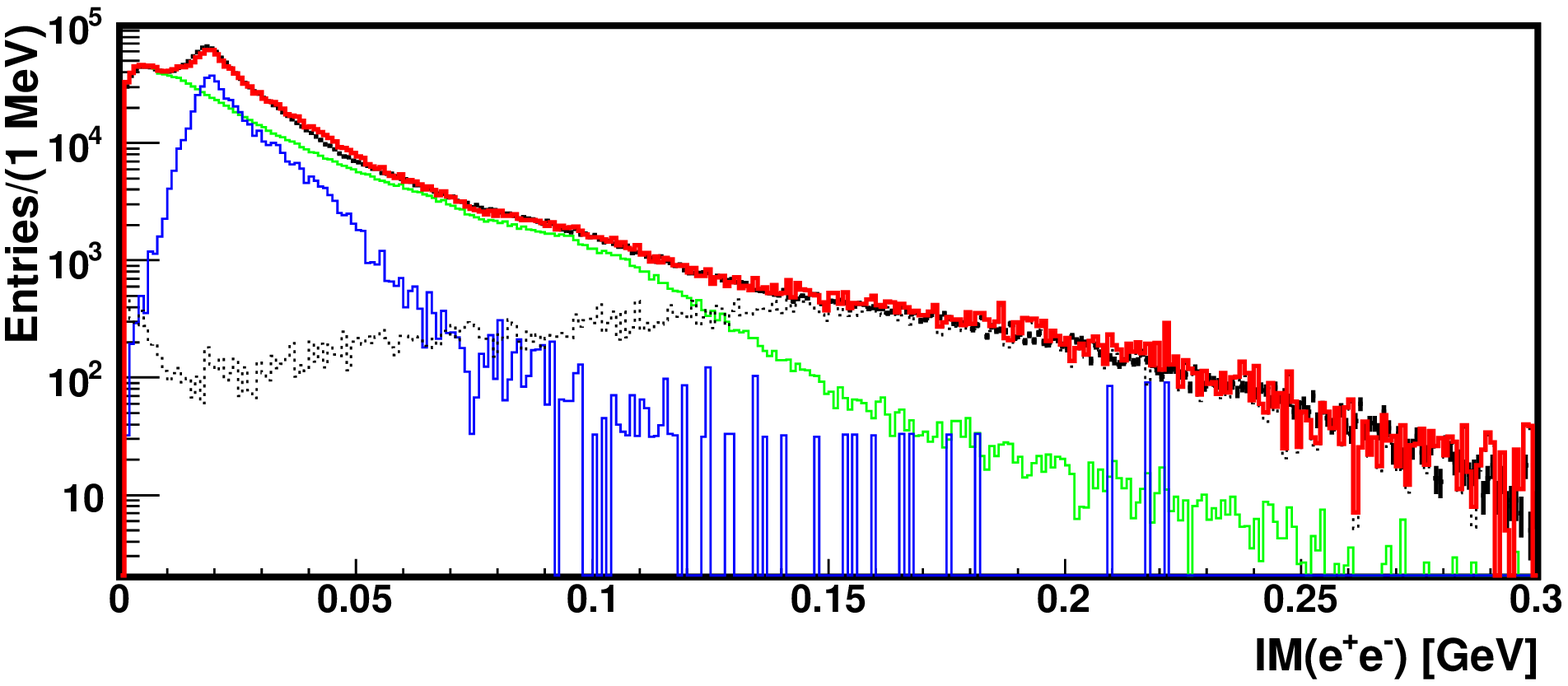}
b)\includegraphics[width=0.98\textwidth]{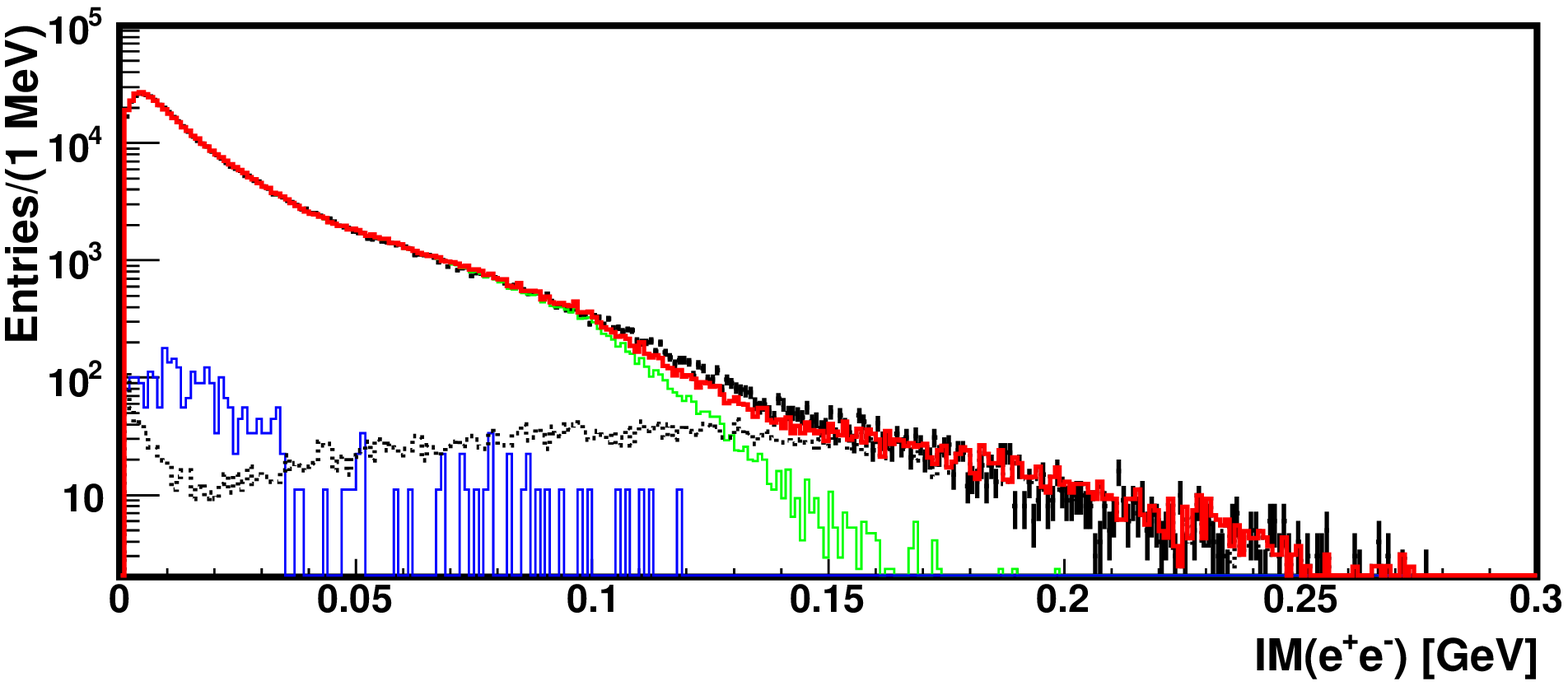}
\caption{\label{fig:imee0}The  reconstructed  $e^+e^-$ invariant  mass
  $q=IM(e^+e^-)$: a)  before and  b) after the  cuts for  reducing the
  conversion background.   The experimental data are  denoted by black
  points.   Results of  simulations  for $\pi^0\to\gamma\gamma$  (blue
  line) and $\pi^0\to e^+e^-\gamma$ (green line) decays are normalized
  according  to  the known  branching  ratios.   The normalization  of
  random coincidences  (dotted line) was fitted in  order to reproduce
  the  $IM(e^+e^-)  >  150$  MeV  range.  The  sum  of  all  simulated
  contributions is given by the red line.  }
\end{figure}

The  first stage  of data  analysis is  to extract  a clean  signal of
$\pi^0\to  e^+e^-\gamma$ decays.   The results  shown in  the previous
section  suggest  that in $pp$ interactions at  550 MeV  electron-positron pairs  come
nearly exclusively from the $\pi^0$ meson decays.  Therefore, in order
to maximize the yield of the $\pi^0\to \gamma e^+e^-$ events we use an
inclusive data sample  requesting events with (i) at  least one proton
identified  in the FD,  (ii) an  $e^+e^-$ pair  identified in  the CD.
There  is no  request  of an  additional  photon cluster  and we  have
included  events  from  both  triggers corresponding  to  phase  space
regions (1) and (2).   The distribution of the reconstructed invariant
mass  of  the  electron-positron  pair, $q=IM(e^+e^-)$,  is  shown  in
Fig.~\ref{fig:imee0}a.  This spectrum is  well described by the sum of
$\pi^0\to   e^+e^-\gamma$  and  $\pi^0\to\gamma\gamma$   (with  photon
conversion).  The data sample contains 1.8\E{6} reconstructed events.

\begin{figure}
\centerline{\includegraphics[width=0.7\textwidth]{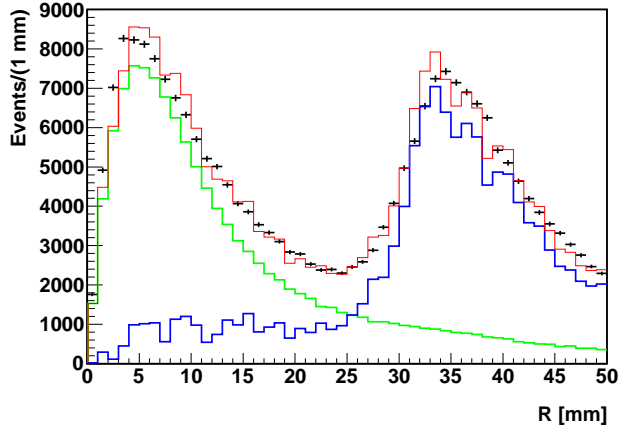}}
\caption{\label{fig:vertex} Distribution  of the distance  $R$ between
  the COSY beam  axis and the reconstructed point  of closest approach
  of  $e^+$  and  $e^-$  tracks: experimental  data  (black  crosses);
  simulations  for $\pi^0\to\gamma\gamma$  (blue line),  the $\pi^0\to
  e^+e^-\gamma$  decay   (green  line),  and   the  sum  of   the  two
  contributions  (red   line).  }
\end{figure}

\begin{figure}
\centerline{\includegraphics[width=0.7\textwidth]{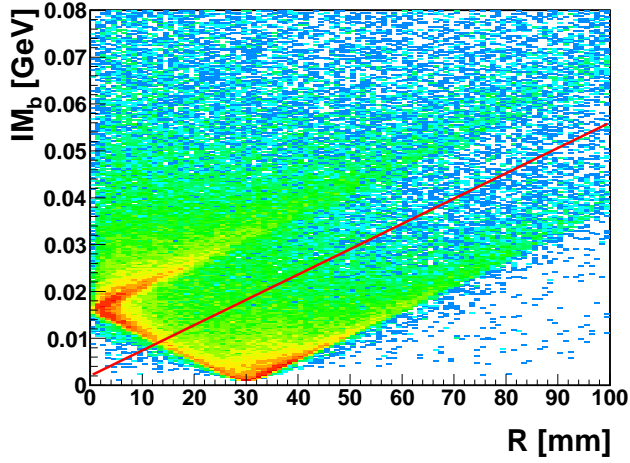}}
\caption{\label{fig:vertex1}   Correlation  between  $R$   and  $IM_b$
  variables for the  experimental data. The selection cut  is shown by
  the diagonal line. The events below the line mainly come from photon
  conversions in the beam pipe. }
\end{figure}

The  $\pi^0\to\gamma\gamma$  events   are  efficiently  removed  by  a
condition  on  the  reconstructed  position of  the  $e^+e^-$  vertex.
Fig.~\ref{fig:vertex}  shows the distance  ($R$) of  the reconstructed
vertex   from  the  COSY   beam  axis.    The  contributions   of  the
$\pi^0\to\gamma\gamma$  and $\pi^0\to e^+e^-\gamma$  decays, simulated
according to  the known branching  ratios, are in very  good agreement
with the observed distribution and  they are well separated.  In order
to  further reduce  the external  conversion background  one  uses the
invariant mass of the $e^+e^-$ calculated from the momentum directions
at the points where the  tracks intersect the beam tube, $IM_b$, shown
in  Fig.~\ref{fig:vertex1}.  The  selection  cut is  performed in  the
$IM_b$ {\it vs.} $R$  plane (Fig.~\ref{fig:vertex1}).  The cut removes
98\%  of   the  $\pi^0\to\gamma\gamma$  events   which  contribute  to
$IM(e^+e^-)$ distribution due to conversion.


The   finally   reconstructed  $dN/dq$   distribution, containing nearly  
 5\E{5} entries, is  shown   in
Fig.~\ref{fig:imee0}b. It is well described by the simulations of the
$\pi^0\to e^+ e^- \gamma$ decay  channel alone with a very small (approx. 3000
events) admixture of background from the $\pi^0\to\gamma\gamma$ decay.
The  data in  this work  represent the  world largest  data  sample of
$\pi^0\to e^+e^-\gamma$ events, which  is almost an order of magnitude
larger than the sample used  for the previously published results from
the SINDRUM experiment \cite{MeijerDrees:1992kd,MeijerDrees:1992qb}.

\begin{figure}
\centerline{\includegraphics[width=0.98\textwidth]{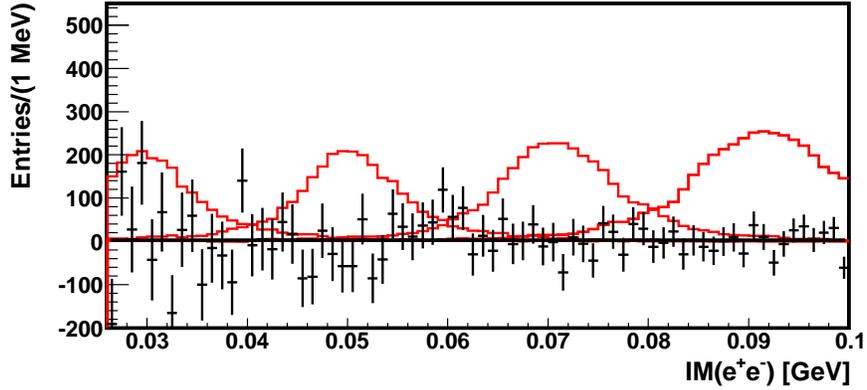}}
\caption[]{\label{fig:imeef}\label{fig:U01}   Difference  between  the
  reconstructed $e^+e^-$  invariant mass  distribution and the  sum of
  all  simulated  contributions (black  points).   The resolution  and
  sensitivity  for a hypothetical decay $U\to e^+e^-$ are  illustrated by  the  superimposed red  histograms.
  They  represent the  signals expected  for the  $\pi^0\to U\gamma\to
  e^+e^-\gamma$ process with $U$ boson  masses of $M_U=$ 30, 50, 70
  and  90 MeV  and $BR(\pi^0\to  U\gamma)=10^{-4}$  (the corresponding
  $\epsilon$ values  are: 0.0077, 0.0088, 0.0113,  and 0.0169
  respectively).  }
\end{figure}

\subsection{Upper limit for  the $BR(\pi^0\to \gamma(U\to e^+e^-))$ }

A   distinctive  feature  of   the  expected   signal  of   the  decay
$\pi^0\to\gamma(U\to e^+e^-)$ (Fig.~1b) is  the appearance of a narrow
peak  (the  width being  given  by  the  detector resolution)  in  the
invariant mass distribution  of the electron positron pair  at the $U$
boson mass.   The electrodynamics process  $\pi^0\to \gamma^*\gamma\to
e^+e^-\gamma$ (Fig.~1a) both represents the irreducible background and
is used for  normalization.  Due to the expected  small decay width of
the $U$ boson the interference  term is negligible and the signal from
the $U$ boson  can be tested by constructing an  incoherent sum of the
two contributions.

The experimental  data are described  well by the simulation  based on
Eqn.~(\ref{matreeg})  alone  as   shown  in  Fig.~\ref{fig:imee0}.   The
difference between reconstructed experimental $q$ distribution and the
sum of all simulated contributions is given in Fig.~\ref{fig:U01}.  
The errors include both statistical uncertainties of the data sample
as well as the systematical ones due to the simulation of the detector response. In
addition   there   are   superimposed   five   example   distributions
corresponding  to the $\pi^0\to  U\gamma\to e^+e^-\gamma$  process for
$U$ boson masses  of 30, 50, 70 and  90 MeV respectively, assuming
$BR(\pi^0\to  U\gamma)=10^{-4}$.    The  plots  illustrate   both  the
resolution and the efficiency expected for the signal. The structure at 
60 MeV is most likely due to a small residual of 
the conversion events which are not yet understood by MC.

For a  given value  of the $U$  boson mass corresponding  to the  range of
the $k^{\rm th}$ bin  of the invariant mass  spectrum ($q_k<M_U<q_k+\Delta q$,
with $\Delta q= 1$ MeV  the width of the histogram bin)  the number of events
in the  $i^{th}$ bin of the  reconstructed electron-positron invariant
mass distribution, $N_i$, can be described in the following form:
\begin{equation}
N_i/N_{Tot}=\frac{1}{\Gamma}\sum_j  S_{ij}  \eta_j \nu_j+S_{ik}
\eta_k\beta\label{eqn:fit}
\end{equation}
The first  term in the Eqn.~(\ref{eqn:fit})  represents the contribution
from   the Dalitz   decay   and   the  second   term   from   the hypothetical
$\pi^0\to\gamma(U\to e^+e^-)$  decay chain. Indices $j$  and $k$ label
the {\em  true, unperturbed}  distributions and $i$  the reconstructed
$q$ histogram.    $N_{Tot}$ is total  number of produced $\pi^0$ mesons, $1/\Gamma$
is the $\pi^0$  life time and $\eta_j$ is  the efficiency. $S_{ij}$ is
the normalized  smearing matrix  (for each $j$:  $\sum_{i} S_{ij}=1$),
$\nu_j$ is the unperturbed $d\Gamma/dq$ distribution for the $\pi^0\to
e^+e^-\gamma$ decay (Eqn.~(\ref{eqn:dgdq2}) and Fig.~\ref{fig:hee2b}a) integrated over bin $j$:
\begin{equation}
\nu_j\equiv         \int_{q_j}^{q_j+\Delta         q}
   \frac{d\Gamma}{dq} dq,
\end{equation}
and $\beta$ is  $BR(\pi^0\to\gamma(U\to e^+e^-))$.  The efficiency and
the smearing  matrix was obtained  from the detector  simulation.  The
$U$  boson decay  mechanism in  diagram  Fig.~\ref{fig:hee2b}b implies
that the efficiencies as a function of $\cos\theta^*$ are identical to
the ones of the $\pi^0\to e^+e^-\gamma$ decay with $q=M_U$.  Note that
for the quoted values of  the branching ratios the intrinsic width 
(\ref{eqn:width})
of the  $U$ boson would be  in the eV  range and thus very  much smaller
than the experimental bin size.

\begin{figure}
\centerline{\includegraphics[width=0.6\textwidth]{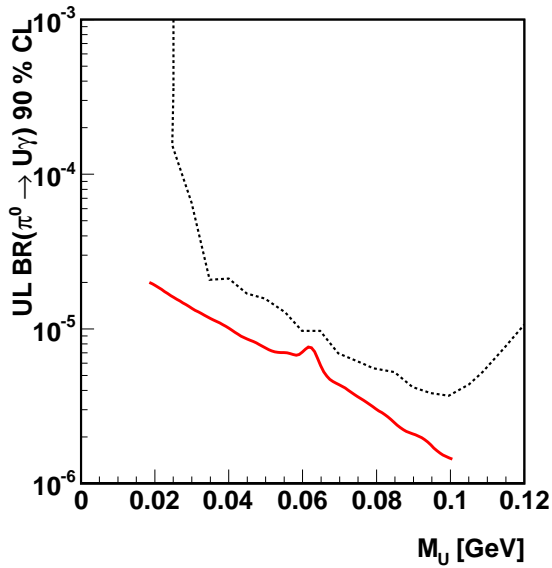}}
\caption[aa]{\label{fig:BRUbos}A  90\%  C.L.    upper  limit  
(smoothed) for  the
  $BR(\pi^0\to\gamma U)$ from this  paper (solid line) compared to
  the result of the SINDRUM experiment \cite{MeijerDrees:1992kd} (dotted
  line).}
\end{figure}

The  upper limits  for the  $U$ boson  branching ratios,  $\beta$,  as a
function of $M_U$  were obtained by repeating for  all bins (index $k$
in Eqn.~(\ref{eqn:fit})),  corresponding to  the $20$ MeV  $< M_U<100$
MeV  range,  the least  square  fits  of  Eqn.~(\ref{eqn:fit}) to  the
experimental $q$  distribution.  
The results of the unconstrained fits yield estimators of $\beta$
values and  their  standard deviations,  which have to a good accuracy  
asymptotic
Gaussian distributions. Finally   we construct the 
upper limits using prescriptions from ref.
\cite{Cowan:2011aa} taking into account the fact that $\beta$ is a non-negative
parameter since the $U$ boson contribution is added incoherently here. 
Fig.~\ref{fig:BRUbos} shows  the 90\%
C.L.  upper  limits  for  the  branching  ratio  of  $\pi^0\to
\gamma(U\to  e^+e^-)$ decay  as a  function  of the  assumed value  of
$M_U$.  This result is compared to that obtained from the SINDRUM data
\cite{MeijerDrees:1992kd}.

\begin{figure}
\centerline{\includegraphics[width=0.6\textwidth]{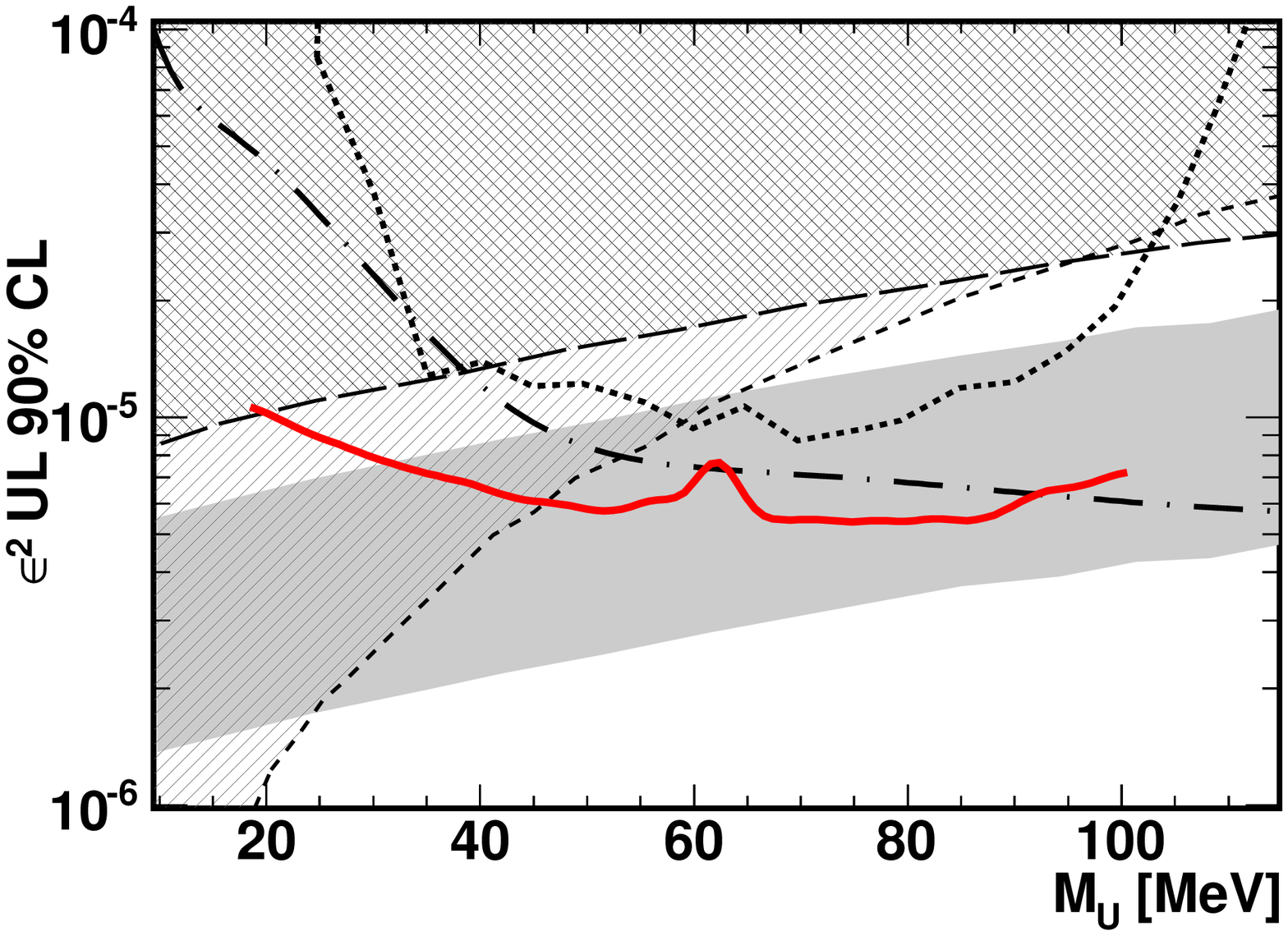}}
\caption[aa]{\label{fig:Ubos}Summary of  the 90\% CL  upper limits for
  the mixing parameter $\epsilon^2$ from WASA-at-COSY (red solid line)
  compared       to        SINDRUM       $\pi^0\to       e^+e^-\gamma$
  \cite{MeijerDrees:1992kd}  (dotted line)  and  recent combined  KLOE
  $\phi\to\eta  e^+e^-$  \cite{Babusci:2012cr}  (dashed dotted)  upper
  limits.   The   long  respectively  short  dashed   lines  (and  the
  corresponding hatched  areas) are the upper limits  derived from the
  muon  and the  electron $g-2$  \cite{Endo:2012hp}.  In  addition the
  gray  area represents the  $\pm 2\sigma$  preferred band  around the
  present value of the muon $g-2$.  }
\end{figure}

The branching  ratio of $\pi^0\to\gamma U$ is  related to $\epsilon^2$
by \cite{Batell:2009di,Reece:2009un}:
\begin{equation}
  \frac{\Gamma(\pi^0\to\gamma U)}{\Gamma(\pi^0\to\gamma\gamma)} = 
2\epsilon^2|F(M_U^2)|^2
\left(1-\frac{M_U^2}{M^2}\right)^3.
  \label{Eq:BR0}
\end{equation}
The resulting upper limits for  the $\epsilon^2$ parameter is shown in
Fig.~\ref{fig:Ubos} and compared with other experiments.

The   recent  limits   for   the  electron   $g-2$   are  taken   from
recent QED calculations
Refs~\cite{Aoyama:2012wj,Endo:2012hp} and a measurement of alpha in atomic physics \cite{Bouchendira:2011aa}.  Our  upper limit improves 
the recent combined KLOE  limits \cite{Babusci:2012cr} at 
low $M_U$.
We use a disparate
experimental  setup and different  meson decay  as source  of $e^+e^-$
pairs.   Together the data  significantly reduce  the parameter  space for
mass and  mixing strength  of a hypothetical  dark photon $U$,  if the
latter is assumed to account for the presently seen deviation between the 
Standard Model prediction and the experimental value of the
muon anomalous magnetic moment.  The experiment presented in the paper
if repeated with  an order of magnitude larger  statistics would cover
the remaining part of this  region of interest. The collected data can
also be used to determine the $\pi^0$ transition form factor.

\section*{Acknowledgments}
This work  was supported in  part by the EU  Integrated Infrastructure
Initiative    HadronPhysics     Project    under    contract    number
RII3-CT-2004-506078;  by   the  European  Commission   under  the  7th
Framework Programme  through the 'Research  Infrastructures' action of
the  'Capacities' Programme,  Call:  FP7-INFRASTRUCTURES-2008-1, Grant
Agreement N.   227431; by the  Polish National Science  Centre through
the     Grants    No.      86/2/N-DFG/07/2011/0    0320/B/H03/2011/40,
2011/01/B/ST2/00431,   2011/03/B/ST2/01847,   0312/B/H03/2011/40
and Foundation for Polish Science.   We
gratefully  acknowledge  the support  given  by  the Swedish  Research
Council,   the  Knut   and  Alice   Wallenberg  Foundation,   and  the
Forschungszentrum J\"ulich FFE Funding  Program of the J\"ulich Center
for Hadron Physics.

We would like to thank Simona Giovannella for providing the KLOE data points.
The authors thank technical staff at Forschungszentrum J\"ulich for support
in preparation of and during the experiment.

This work is part of the PhD Thesis of C.-O.~Gullstr{\"o}m.

\bibliographystyle{h-elsevier}
\bibliography{pap}

\end{document}

%% file: author1.tex
\author[IKPUU]{The WASA-at-COSY Collaboration\\[2ex] P.~Adlarson}
\author[ASWarsN]{W.~Augustyniak}
\author[IPJ]{W.~Bardan}
\author[PITue,Kepler]{M.~Bashkanov}
\author[MS]{F.S.~Bergmann}
\author[ASWarsH]{M.~Ber{\l}owski}
\author[IITB]{H.~Bhatt}
\author[Budker]{A.~Bondar}
\author[IKPJ,JCHP]{M.~B\"uscher}
\author[IKPUU]{H.~Cal\'{e}n}
% ...

\address[IKPUU]{Division of Nuclear Physics, Department of Physics and 
 Astronomy, Uppsala University, Box 516, 75120 Uppsala, Sweden}
\address[ASWarsN]{Department of Nuclear Physics, National Centre for Nuclear 
 Research, ul.\ Hoza~69, 00-681, Warsaw, Poland}
\address[IPJ]{Institute of Physics, Jagiellonian University, ul.\ Reymonta~4, 
 30-059 Krak\'{o}w, Poland}
\address[PITue]{Physikalisches Institut, Eberhard--Karls--Universit\"at 
 T\"ubingen, Auf der Morgenstelle~14, 72076 T\"ubingen, Germany}
\address[Kepler]{Kepler Center for Astro and Particle Physics, Eberhard Karls 
 University T\"ubingen, Auf der Morgenstelle~14, 72076 T\"ubingen, Germany}
\address[MS]{Institut f\"ur Kernphysik, Westf\"alische Wilhelms--Universit\"at 
 M\"unster, Wilhelm--Klemm--Str.~9, 48149 M\"unster, Germany}
\address[ASWarsH]{High Energy Physics Department, National Centre for Nuclear 
 Research, ul.\ Hoza~69, 00-681, Warsaw, Poland}
\address[IITB]{Department of Physics, Indian Institute of Technology Bombay, 
 Powai, Mumbai--400076, Maharashtra, India}
\address[Budker]{Budker Institute of Nuclear Physics of SB RAS, Academician
Lavrentyev~11, Novosibirsk, 630090, Russia}
\address[IKPJ]{Institut f\"ur Kernphysik, Forschungszentrum J\"ulich, 52425 
 J\"ulich, Germany}
\address[JCHP]{J\"ulich Center for Hadron Physics, Forschungszentrum J\"ulich, 
 52425 J\"ulich, Germany}
% ...

\fntext[fnde]{present address: Department of Physics and Astrophysics, 
 University of Delhi, Delhi--110007, India}
\fntext[fnsu]{present address: Department of Physics, Stockholm University, 
 Roslagstullsbacken~21, AlbaNova, 10691 Stockholm, Sweden}
\fntext[fnmz]{present address: Institut f\"ur Kernphysik, Johannes 
 Gutenberg--Universit\"at Mainz, Johann--Joachim--Becher Weg~45, 55128 Mainz, 
 Germany}
\fntext[fnuc]{present address: Department of Physics and Astronomy, University 
 of California, Los Angeles, California--90045, U.S.A.}
\fntext[fnbe]{present address: Albert Einstein Center for Fundamental Physics,
 Fachbereich Physik und Astronomie, Universit\"at Bern, Sidlerstr.~5, 
 3012 Bern, Switzerland}

\cortext[coau]{Corresponding author }